\documentclass[aps,prl,twocolumn,groupedaddress,floatfix]{revtex4}

\usepackage{graphicx}
\usepackage{subfigure}
\usepackage{epsfig}
\usepackage{amsmath}
\usepackage{bbm}

\def\be{\begin{equation}}
\def\ee{\end{equation}}
\def\bea{\begin{eqnarray}}
\def\eea{\end{eqnarray}}

\begin{document}


\title{Quantum Hall effect in carbon nanotubes}


\author{E. Perfetto$^{1,3}$, J. Gonz{\'a}lez$^1$, F. Guinea$^2$,
      S. Bellucci$^3$ and P. Onorato$^{3,4}$}
\affiliation{$^1$ Instituto de Estructura de la Materia.
        Consejo Superior de Investigaciones Cient{\'\i}ficas.
        Serrano 123, 28006 Madrid. Spain.  \\
             $^2$Instituto de Ciencia de Materiales.
        Consejo Superior de Investigaciones Cient{\'\i}ficas.
        Cantoblanco. 28049 Madrid. Spain.   \\
             $^3$INFN, Laboratori Nazionali di Frascati,
              P.O. Box 13, 00044 Frascati, Italy. \\
             $^4$Dipartimento di Scienze Fisiche,
        Universit\`{a} di Roma Tre, Via della Vasca Navale 84,
                           00146 Roma, Italy.}

\date{\today}

\begin{abstract}
We investigate the effects of a transverse magnetic field on the 
transport properties of carbon nanotubes, making use of a 
long-wavelength description in terms of Dirac fermion fields.
For values of the magnetic length smaller than the nanotube radius,
we observe that the electronic states organize into incipient Landau
subbands, with a highly degenerate level at zero energy. We show
that only the states in dispersive branches, localized at the 
flanks of the nanotube, are able to transport current in the longitudinal
direction. This is at the origin of the quantization of the Hall 
conductivity, that turns out to be given by even multiples of $2 e^2 /h$.
We also analyze the effects of the electron-electron interaction, showing
that the magnetic field induces a suppression of the electronic correlations,
reflected in particular in the enhancement of the tunneling density of
states near the Fermi level.

\end{abstract}

\maketitle

Two-dimensional carbon compounds with $sp^2$ bonding have attracted 
recently much attention, due to the experimental observation of a number
of novel electronic properties. It has been possible to measure the 
transport properties of a single layer of graphite (so-called graphene), 
providing evidence that the quasiparticles have a conical dispersion
around discrete Fermi points\cite{novo,zhang}. 
This seems to be at the origin of 
remarkable features of the resistivity as well as of the Hall 
conductivity. Carbon nanotubes can be also considered as the result of
wrapping up the graphene sheet, leading to systems where the transport
seems to be ballistic under certain conditions\cite{heer}, as a 
consequence of the suppression of the scattering between different 
low-energy subbands\cite{egger,kane}.

The metallic carbon nanotubes and the graphene sheet have in common
that their low-energy electronic dispersion is governed by a massless 
Dirac equation, around each of the two Fermi points of the undoped 
systems\cite{mele1,nos,mele2}. 
The low-energy electronic states can be encoded into a couple of Dirac 
fermion fields, and the appearance of an additional pseudo-spin quantum 
number gives rise to important effects in the spectrum. This new degree 
of freedom has allowed us  to understand, for instance, the degeneracy of 
the molecular orbitals in the fullerenes\cite{prl}, the quantization 
rule of the Hall conductivity in graphene\cite{graph1,graph2}, or the 
properties of the polarizability in carbon nanotubes\cite{lev}.

In this paper we investigate the effects of a transverse magnetic field
on the transport properties of the carbon nanotubes, making  
use of the description of the electronic states in terms of Dirac fermion
fields. While it is well-known the effect of a magnetic field 
parallel to the tube axis, the transport properties of carbon nanotubes under
a transverse magnetic field are less understood\cite{dress}. 
We will see that carbon nanotubes of sufficiently large radius 
may also have a quantum Hall regime, with a quantized Hall conductivity
$\sigma_{xy}$. In the case of graphene, it has been shown that 
$\sigma_{xy}$ has plateaus at odd multiples of $2 e^2 /h $\cite{novo,zhang}, 
as a consequence of the peculiar Dirac spectrum\cite{graph1,graph2}. We will 
find that the different topology of the carbon nanotubes leads instead to a 
quantization in even steps of the quantity $2 e^2 /h $.

For the Hall regime to arise,
the diameter of the nanotubes has to be larger than the the magnetic 
length $l = \sqrt{\hbar c /e B}$, since this is the typical size of the 
localized states in Landau levels. Suitable conditions can be already 
found in thick multi-walled nanotubes for magnetic fields
$B \gtrsim 1 \; {\rm T}$, which correspond to magnetic lengths 
$l \lesssim 30 \; {\rm nm}$. We will see that, for $l < R$, the 
eigenstates of the carbon nanotube organize into incipient Landau 
subbands, with a highly degenerate level at zero energy. The branches
with linear dispersion correspond to states 
localized at the flanks of the nanotube, carrying quantized currents 
which are responsible for the conductivity along the longitudinal 
dimension of the nanotube.

In order to establish a relation with the quantum Hall effect in 
graphene, it is convenient to set up an approach focusing on the features
of the states over distances much larger than the $C$-$C$ distance $a$.
The low-energy band structure can be obtained in graphene by taking a 
continuum limit in which the momenta are much smaller than the inverse 
lattice spacing $1/a$ \cite{mele1,nos,mele2}. In the case of carbon 
nanotubes under magnetic field, 
a sensible continuum limit requires also that $l \gg a$, so that lattice 
effects can be disregarded. In that limit, we obtain a simple field theory 
of Dirac spinors coupled to the magnetic field, allowing us to investigate 
different features of the quantum Hall effect in the tubular geometry.

We illustrate the long-wavelength limit in the case of zig-zag 
nanotubes. These have a unit cell with length $3a$, containing four 
transverse arrays of $N$ carbon atoms at different longitudinal positions
$x_i, i = 1, \ldots 4$. We introduce the Fourier transform of the electron 
operator $c(x_i,n)$ with respect to the position $n = 1,2, \ldots N$ in the 
transverse section
\begin{equation}
c (x_i,n) \sim \sum_p d_p (l;i) e^{i 2\pi np/N}
\end{equation}
where $l \in Z$ runs over the different cells.
The index $p$ labels the different 1D subbands, 
$p = 0, 1, \ldots N-1$. Their dispersion can be obtained from
the diagonalization of a 1D system with four orbitals per unit cell.
This leads in general to massive subbands with parabolic dispersion,
with a gap $2 \Delta_p = 2 t | 1 - 2 \cos (  \pi p / N ) |$, where 
$t \approx 3$ eV is the tight-binding element between two 
nearest-neighbor carbon atoms. The
dispersive branches can be decoupled from the high-energy branches
that appear near the top of the spectrum. It turns out that the
low-energy dispersion corresponds to a reduced 2-component
spinor, with a hamiltonian depending on the subband index $p$ and
the longitudinal momentum $k$
\begin{equation}
  {\cal H}_{p,p'}^0    =
  \delta_{p, p'}  \left(
\begin{array}{cc}
 \hbar v_F k     &    \Delta_p     \\
\Delta_p      &    - \hbar v_F k
\end{array}         \right)
\end{equation}
where the Fermi velocity is $v_F = 3ta/2 \hbar $.

In the presence of a transverse magnetic
field, the electron field picks up in general a factor $e^{i\phi}$ when 
transported between nearest-neighbor sites of the carbon lattice, with
$\phi \propto a (e/\hbar c) BR \sin (2\pi n/N)$ \cite{dress}. In the 
continuum limit, characterized by $(e/\hbar c) BRa \ll 1$, we can linearize 
in the strength of the magnetic field. This
introduces an interaction which is nondiagonal in the space of the
different subbands, and that can be represented by the operator
\begin{equation}
\Delta {\cal H}_{p,p'}    =
  \delta_{p', p \pm 1}  \times \left( \pm i v_F \frac{e}{c} \frac{BR}{2} \right)
  \times \left(
\begin{array}{cc}
   1     &    0    \\
   0    &    - 1
\end{array}         \right)
\end{equation}

The total hamiltonian ${\cal H} = {\cal H}^0 + \Delta {\cal H}$
can be more easily expressed when acting on the Dirac
spinor $\Psi (k;\theta )$ depending on the angular variable
$\theta $ around the tubule. In this basis there are in fact two
different sectors describing states about the two angular momenta
$\pm P$ with vanishing gap, $\Delta_{\pm P} \approx 0$.
The hamiltonian is in either sector
\begin{equation}
{\cal H} =  \left(
\begin{array}{cc}
 \hbar v_F k  +  v_F \frac{eBR}{c} \sin (\theta )  
                      &  - i(\hbar v_F /a ) \partial_{\theta }  \\
 - i(\hbar v_F /a ) \partial_{\theta }  
                &  - \hbar v_F k  -  v_F \frac{eBR}{c} \sin (\theta )
\end{array}
   \right)
\label{dirac}
\end{equation}
where the periodic modulation matches with the orientation of a
magnetic field normal to the nanotube surface at $\theta = 0$.
This hamiltonian resembles that for a nanotube in a transverse electric
field\cite{lev}, although it poses the additional difficulty that the
perturbation $\Delta {\cal H}$ hybridizes positive and negative energy
eigenstates of ${\cal H}^0$.
Expression (\ref{dirac}) corresponds actually to the Dirac hamiltonian
with the usual prescription for the coupling to the vector potential.
We have checked that, as expected, the eigenstates of (\ref{dirac}) provide
a good approximation to the low-energy band structure of the carbon
nanotubes for $a R/l^2 \ll 1$.

\begin{figure}
\begin{center}
\mbox{\epsfxsize 7.5cm \epsfbox{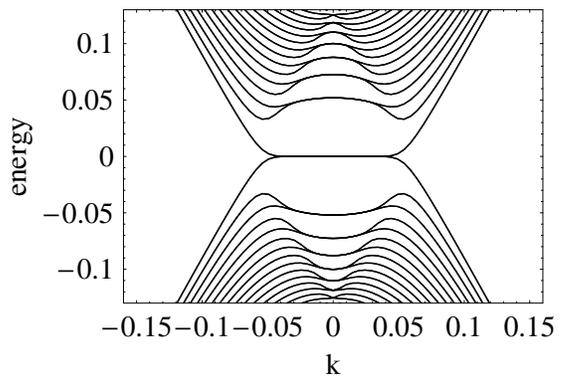}}
\end{center}
\caption{Band structure of a zig-zag nanotube in transverse magnetic 
field, for a radius $R \approx 20$ nm and field strength $B=20$ T. This 
choice corresponds to $aR/l^{2} \approx 0.1$ and $R/l \approx 3.5$. 
Energy is in units of $t$ and momentum is in units of \AA$^{-1}$.}
\label{one}
\end{figure}

It can be shown that, for intermediate values of the magnetic length
$l \sim R$, the spectrum starts to develop a flat Landau level at
zero energy, similar to that in the case of graphene. According
to the mentioned doubling of the subbands, there is a four-fold degeneracy 
for each momentum $k$ in the flat level. This extends into dispersive 
branches with particle- and hole-like character, at each side in momentum 
space. A typical band structure is represented in Fig. \ref{one}.
It can be checked that the energy levels at $k = 0$ follow the 
quantization rule $\varepsilon_n \propto \sqrt{n}$, which is peculiar of 
graphene\cite{mac}.
The plot corresponds to a zig-zag nanotube, but it can be shown that
the shape of the band structure remains the same for other geometries, 
with the two valleys at zero energy (that appear superposed in Fig. 
\ref{one}) expanding in general around the two Fermi points of the 
system at $B =0$.

A remarkable point is that each eigenfunction of (\ref{dirac}) is
in general localized around a certain value of the angular variable
$\theta $. The zero-energy states at $k = 0$, for instance, have
gaussian wave functions localized at $\theta = 0$
or $\theta = \pi$. For positive (negative) longitudinal momentum,
the zero-energy states are localized at angles between 0 and
$\pi /2 $ ($-\pi /2 $), or between $\pi $ and $\pi /2 $ ($-\pi /2 $),
depending on the subband chosen. Quite interestingly, the states
in the dispersive branches have gaussian wave functions centered
about $\pi /2 $ (for a right branch) or $-\pi /2 $ (for a left
branch).

\begin{figure}
\begin{center}
\mbox{\epsfxsize 7.5cm \epsfbox{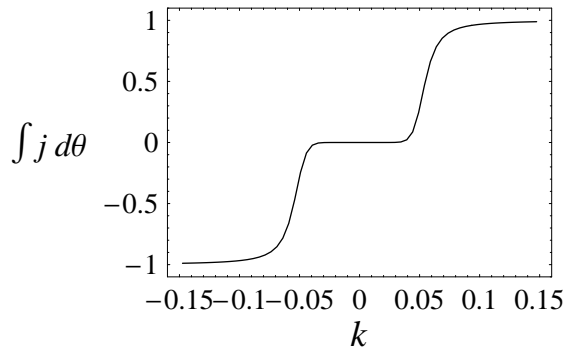}}
\end{center}
\caption{Plot of the integral of the current $j$ over the angular 
variable $\theta $, for states in the lowest Landau subband labeled by 
the longitudinal momentum $k$ (in units of  \AA$^{-1}$).
The parameters are the same as in Fig. 1.}
\label{two}
\end{figure}

The localization of the states in the dispersive branches at the
flanks of the tubule suggests that, despite having no boundary,
the carbon nanotube may support edge states in similar fashion as
in systems with planar geometry. To check this fact, one may compute
the current flowing in the longitidunal direction for the different
states. For the Dirac spinors, the definition of the current follows
from the continuity equation
\begin{equation}
\partial_t (\Psi_R^{+} \Psi_R + \Psi_L^{+} \Psi_L )
   = v_F \partial_x (\Psi_R^{+} \Psi_R - \Psi_L^{+} \Psi_L )  
\end{equation}
where $\Psi_R $ and $\Psi_L $ are the two spinor components, in
the basis used to write the Dirac hamiltonian (\ref{dirac}). We have
computed the integral over $\theta $ of the current
\begin{equation}
j = \Psi_R^{+} \Psi_R - \Psi_L^{+} \Psi_L 
\end{equation}
for the different eigenstates. The results for the lowest
energy subbands are represented in Fig. \ref{two}. It turns out that,
in general, the states corresponding to the flat part of the
Landau level do not carry any current in the longitudinal direction,
while the states in the dispersive branches saturate quickly the
unit of current as the dispersion approaches a constant slope.

The quantization of the current for the states in the dispersive
branches opens the possibility to observe the quantization
of the Hall conductivity in thick carbon nanotubes. In general,
the quantization of the current is more accurate for smaller
curvature of the dispersive branches. It happens moreover that, 
when the Fermi level crosses one of the bumps with parabolic
dispersion shown in Fig. \ref{one}, the two contributions to the
current from the respective Fermi points go in the opposite direction
and tend to cancel one each other. Envisaging an experiment where
a potential difference is applied between the two flanks of a
thick nanotube, we obtain that the current in the longitudinal
direction is given approximately by the excess (or defect) of
filled states in the right dispersive branches, with respect to
those in the left dispersive branches. Making the parallel of the
arguments applied for planar geometries\cite{halp}, we conclude that 
the Hall conductivity $\sigma_{xy}$ must follow an approximate 
quantization rule, with a prefactor given by the spin degeneracy and 
the doubling of the subbands shown in Fig. \ref{one}:
\begin{equation}
\sigma_{xy} = 4 \frac{e^2}{h} n
\end{equation}

Opposite to what happens in the case of graphene, we observe
that the Hall conductivity is quantized in even steps of
$2 e^2 /h $. It can be shown that this feature is a consequence of
the vanishing net flux traversing the nanotube surface. We have checked 
that cutting the nanotube along the longitudinal direction at $\theta = \pi/2 $ 
does not lead to an appreciable change in the band structure, while cutting 
it at $\theta = 0 $ or $\pi $ produces a sudden distortion of the subbands. 
In the latter case, states that were before at zero energy form two subbands 
dispersing towards higher energies, and two other dispersing downwards. In 
these conditions, only one valley is left at zero energy, evidencing the 
transition to the odd-integer quantization of the Hall conductivity of 
graphene.

The coexistence of a magnetic field and a Hall voltage, $V_H$, leads to an
interesting realization of the setup discussed in Ref. \cite{LSB06}. The Hall
voltage leads to an electric field at the top and bottom regions of the
nanotube, ${\cal E} \approx V_H / R$. Due to the 1D nature of
the nanotube, this field will not be screened. 
The effect of this field on the Landau levels can be analyzed by noticing 
that it can be made to vanish by a suitable Lorentz transformation, leading 
to a new effective hamiltonian with a reduced magnetic field\cite{LSB06}. 
Hence, one obtains that the new Landau levels are determined
by the magnetic length $l' = l ( 1 - \beta^2 )^{-1/4}$, where $\beta = {\cal
  E} / [ ( v_F / c ) B ]$. The system undergoes a phase transition,
similar to dielectric breakdown\cite{LSB06}, for $\beta=1$. In the case of a
nanotube, this transition takes place when $( V_H^c / R ) / [ ( v_F / c ) B ] =
1$. This corresponds to a current flowing along the nanotube:
\begin{equation}
I_c = \sigma_{xy} V_H^c \sim 4n \frac{e v_F R}{l^2} 
\end{equation}
The transition at higher bias currents leads to a significant change in the
electronic wavefunction, and to the suppression of the gaps between Landau
levels. It seems likely that, beyond this transition, the Hall conductance
will no longer be quantized.

The existence of extended states along the flanks of the nanotube
gives also the clue to understand the effects of the electron-electron
interaction in the presence of the magnetic field. At $B = 0$, the
transport properties are dictated by the so-called Luttinger liquid
behavior, which is a reflection of the repulsive interaction in the
nanotubes\cite{egger,kane}. When the magnetic field is switched on, 
however, the Coulomb interaction renormalizes in a different manner 
the Fermi velocity as well as the compressibility for the extended 
states near the Fermi level, with the consequent modification of the 
transport properties. 

\begin{figure}
\begin{center}
\mbox{\epsfxsize 7.5cm \epsfbox{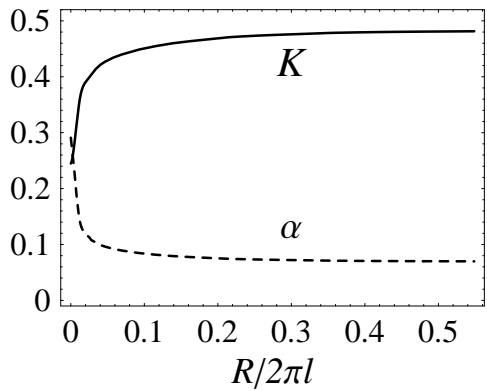}}
\end{center}
\caption{Plot of the Luttinger liquid parameter $K$ (solid line) and 
of the critical exponent $\alpha$ (dashed line) as a function of 
$R / 2 \pi l$. Here the radius is fixed at $R \approx 20$ nm and $B$ is 
ranging between 0 T ($l=\infty $) and 20 T ($l \approx 6$ nm).} 
\label{three}
\end{figure}

Focusing on the case where the Fermi level is right above (or below)
the plateau at zero energy, the extended states must correspond to the
outermost dispersive branches in the band structure of Fig.
\ref{one}. In the 1D notation of the different interaction channels,
the Fermi velocity is renormalized by the so-called $g_4$ coupling,
given by the matrix element of the Coulomb interaction for electrons 
near the same Fermi point. This quantity can be computed using the
eigenfunctions of the Dirac hamiltonian (\ref{dirac}). At the flanks
of the nanotube, their wave functions have gaussian shape, with a width
that is proportional to $\sqrt{l}$. Thus, 
the coupling $g_4$ gets larger values for increasing magnetic field, 
as a consequence of the strong Coulomb repulsion between the 
currents localized at a given flank of the nanotube. The matrix element
computed instead for electrons near the Fermi level with opposite 
longitudinal momenta gives the so-called $g_2$ coupling. 
The weak Coulomb repulsion between currents localized at antipodal
points in the nanotube leads to relative small values of $g_2$.
This enters in the expression of the Luttinger liquid parameter
\begin{equation}
K = \sqrt{  \frac{v_F + 2(g_4 - g_2)/ \pi \hbar}
                           {v_F + 2(g_4 + g_2)/ \pi \hbar }  }
\end{equation}
which governs the low-energy transport properties. The variation of 
$K$ upon switching on the magnetic field, represented in Fig. \ref{three},
translates into a sharp decrease of the 
exponent $\alpha $ for the the power-law behavior of the tunneling 
density of states, according to the relation
$\alpha = (K + 1/K - 2)/8$ \cite{egger,kane}. 
The behavior of the exponent, plotted in Fig. \ref{three}, provides a 
signature of the suppression of the electronic correlations in 
the presence of a transverse magnetic field, which seems to have been 
observed in the measures of $\alpha $ in multi-walled 
nanotubes\cite{kanda}.

To summarize, we have shown that, for thick carbon nanotubes in a
transverse magnetic field, the transport properties are governed by
the states localized at the flanks of the nanotube, which carry 
quantized currents in the longitudinal direction. By placing a 
potential difference at opposite sides of a nanotube section,
it should be possible to observe steps in the Hall conductivity at
even multiples of $2 e^2 /h $. Furthermore, the small overlap between
states with currents flowing in opposite direction must lead to 
the absence of significant backscattering interactions. This means that
the effects of the $e$-$e$ interactions have to reflect
in a simple renormalization of physical parameters, and that there
should be good perspectives to observe a robust chiral liquid at the 
flanks of the nanotube, by means for instance of scanning tunneling 
microscopy.

The financial support of the Ministerio
de Educaci\'on y Ciencia (Spain) through grants
FIS2005-05478-C02-01/02  is gratefully acknowledged.
E. P. was also supported by INFN grant 10068.


\end{document}